\documentstyle[floats,twocolumn,pra,aps,psfig]{revtex}

\begin{document}
\title{Thermoelastic effects at low temperatures and\\
quantum limits in displacement measurements}
\author{M.\ Cerdonio\thanks{%
visiting at Ecole Normale Sup\'{e}rieure, Laboratoire Kastler Brossel} and
L.\ Conti}
\address{INFN\ Section of Padova and Department of Physics, University of Padova,
Italy}
\author{A.\ Heidmann and M.\ Pinard}
\address{Laboratoire Kastler Brossel\thanks{%
Laboratoire de l'Universit\'{e} Pierre et Marie Curie et de l'Ecole Normale
Sup\'{e}rieure associ\'{e} au Centre National de la Recherche Scientifique},
4 place Jussieu, F75252\ Paris, France}
\date{September 29, 2000}
\maketitle

\begin{abstract}
The displacement fluctuations of mirrors in optomechanical devices, induced
via thermal expansion by temperature fluctuations due either to
thermodynamic fluctuations or to fluctuations in the photon absorption, can
be made smaller than quantum fluctuations, at the low temperatures, high
reflectivities and high light powers needed to readout displacements at the
standard quantum limit. The result is relevant for the design of quantum
limited gravitational-wave detectors, both ''interferometers'' and ''bars'',
and for experiments to study directly mechanical motion in the quantum
regime.\bigskip 
\end{abstract}

\bigskip

\section{Introduction}

In a recent paper Braginsky {\it et al}\cite{Braginsky}, henceforth BGV,
considered the noise in interferometric gravitational-wave detectors due to
thermoelastic fluctuations of the mirrors attached to the test masses of the
interferometer. These thermoelastic fluctuations have contributions from two
independent processes, both acting via the thermal expansivity of the mirror
substrate material.\ The first one is the {\it thermodynamic} fluctuations
in temperature of the body of the mirror substrate (these, in the
approximation of small thermal expansion, are independent from the
thermodynamic fluctuations in volume, which are responsible for the well
studied {\it thermal} or {\it brownian} noise\cite{Saulson}).\ The second
one is the {\it photothermal} temperature fluctuations due to the fact that
the number of photons absorbed by the mirror fluctuate.

BGV results for the thermodynamic noise, obtained for half-infinite mirrors,
have been extended to the case of finite size mirrors\cite{Liu}, with
particular reference to the design of advanced interferometric
gravitational-wave detectors, such as LIGOII\cite{LigoII}. In both cases the
calculations are concerned with mirrors at room temperature, made of
materials well in use for mirrors substrates as fused silica and sapphire,
with km long Fabry-Perot cavities, which are characterized by laser beam
spots of size $r_{0}\simeq 1~cm$ and comparatively low finesse ${\cal F}%
\simeq 100$, and with characteristic frequencies $f\simeq 100~Hz$, for the
mechanical motion to be monitored optically.

There are a number of situations, at variance with the above, which are of
interest for optomechanical devices. In such situations one or both of the
thermoelastic fluctuations effects may be of concern, when one would like to
reach, in the measurements of small displacement, the so called {\it %
Standard Quantum Limit} (SQL)\cite{Caves,BraginskyBook}. Already for LIGOII,
BGV seems to discourage, in favour of fused silica, the use of sapphire,
which on the other hand may be the material considered for {\it cold mirrors}
in connection with advanced configurations of interferometric
gravitational-wave detectors, under study as LCGT\cite{LCGT}, LIGOIII \cite
{LigoII} and EURO\cite{Giazotto}.

The BGV effects would be of concern for very sensitive displacement sensors
based on high-finesse Fabry-Perot cavities, to be used in connection with
bar detectors of gravitational waves as dual cavity transducers\cite{Conti},
or to study the quantum effects of radiation pressure\cite
{Cohadon,Pinard99,Heidmann97}.\ In both cases the cavities are much shorter
(less than a few centimeter) than in a gravitational-wave interferometer,
the beam spots are smaller ($r_{0}\simeq 10^{-2}~cm$), finesses much larger (%
${\cal F}\gtrsim 10^{5}$) and temperatures as low as $T\lesssim 1~K$. It may
appear from BGV results that the thermoelastic effects would generate
particularly large effects, inasmuch the volume involved in the fluctuation
processes would be correspondingly smaller.

For these reasons, it is of interest to explore what would be the behaviour
of both thermoelastic effects in the low temperatures and small beam spot
regimes, where some BGV assumptions break down.\ In particular, the heat
diffusion length $l_{t}$ depends on the temperature and can become larger
than the laser beam spot dimension $r_{0}$, so that the adiabatic
approximation is no longer valid.

In Section \ref{HeatPropag} we give the essentials of the regime of phonons
and heat propagation, which establishes at low enough temperatures, and we
evaluate the thermoelastic noises with a simple calculation, in relation to
the beam spot size and to the frequencies at which the optomechanical device
is most sensitive.

In Sections \ref{Thermodynamic} and \ref{PhotoThermal} we give an exact
calculation of both thermoelastic effects in the whole region of interest,
that is for any value of the ratio $l_{t}/r_{0}$. The results, under the
assumptions of low temperature regime of Section \ref{HeatPropag}, would
directly apply to actual mirrors for the quoted optomechanical devices. We
also relate in a general way the photothermal noise to the displacement
noise induced on the mirror by the quantum fluctuations of radiation
pressure in the cavity.

In Section \ref{Discussion} we discuss limitations and relevance of our
approach in the design of the SQL optomechanical displacement devices.

\section{Heat propagation at low temperatures}

\label{HeatPropag}Let us assume the optomechanical device to work in some
frequency range centered around a frequency $f$ and let us discuss the
photothermal effect. We revisit the calculation of BGV in the following way,
so to use it to see the regime which sets up at low temperatures.

The multilayers coating of the mirror absorbs a small fraction of the light
power and this induces an inhomogeneous increase of the temperature of the
bulk. The absorbed power is a Poisson distributed random variable (the
statistics of the absorbed photon will be discussed in more details in
section \ref{PhotoThermal}), and these fluctuations lead to thermal
fluctuations in the bulk of the mirror.\ They are consequently responsible
for fluctuations of the position of the reflecting face of the mirror, via
the thermal expansion of the mirror material.

The r.m.s. displacement noise of the mirror end face $\Delta z=z\alpha
\Delta T$ is found by evaluating the r.m.s. fluctuation in temperature $%
\Delta T$, in a volume $V$ of the mirror of thickness $z$, linear thermal
expansion coefficient $\alpha (T)$ and specific thermal capacity $C(T)$, as
the absorbed photon flux $n$ fluctuates, 
\begin{equation}
\Delta z=z\alpha \frac{\hbar \omega _{0}\Delta n}{\rho CV},
\label{Eq_Deltax}
\end{equation}
where $\hbar \omega _{0}$ is the energy per photon, $\Delta n=\sqrt{%
\overline{n}/f}$ is the r.m.s.\ poissonian fluctuation of the number of
photons absorbed over the time $1/f$ ($\overline{n}$ is the average absorbed
photon flux), and $\rho $ is the density of the mirror material (axis $z$ is
taken normal to the plane face of the mirror).

At room temperature and for large beam spots, BGV conditions apply: the
phonon mean free path and relaxation times are very small respectively in
comparison to the mirror coating thickness (where the photons create the
phonons in the absorption process), and in comparison with the
characteristic time $1/f$. The thermal diffusion length at frequency $f$ is
given by 
\begin{equation}
l_{t}=\sqrt{\frac{\kappa }{\rho Cf}},  \label{Eq_lt}
\end{equation}
where $\kappa $ is the thermal conductivity. For a frequency $f$ around $%
100~Hz$, $l_{t}$ is on one hand larger than the coating thickness $z_{c}$,
and on the other hand much smaller than the beam spot radius $r_{0}$, 
\begin{equation}
z_{c}<l_{t}<r_{0}.  \label{Eq_ltCond}
\end{equation}
This is the basic BGV approximation, which gives that the volume involved in
the fluctuating thermal expansion effect is the fraction of mirror substrate 
$V\simeq l_{t}r_{0}^{2}$ and thus one has $z\simeq l_{t}$ in Eq. (\ref
{Eq_Deltax}).

This argument reproduces the essential features of BGV spectral density $%
S_{z}\left[ f\right] $ of photothermal displacement noise, as one may write
around the frequency $f$, 
\begin{equation}
S_{z}\left[ f\right] \simeq \frac{\Delta z^{2}}{f}\simeq \left( \frac{\alpha 
}{\rho Cr_{0}^{2}}\right) ^{2}\frac{S_{abs}}{f^{2}},  \label{Eq_Sxxp}
\end{equation}
where $S_{abs}=\hbar \omega _{0}W_{abs}$ is the spectral power noise of the
absorbed light, with $W_{abs}=\hbar \omega _{0}\overline{n}$ the average
absorbed light power. In fact we see that Eq. (\ref{Eq_Sxxp}) is the same
final BGV relation (Eq.\ 8 of \cite{Braginsky}), apart from a term with the
Poisson ratio of the mirror material and numerical factors.

The condition (\ref{Eq_ltCond}) may break down for small beam spot radius $%
r_{0}$ or for low temperature $T$, as the thermal diffusion length gets
longer, either in the mirror substrate or in the mirror coating or in both.

For mirrors substrates of crystalline materials, as specifically sapphire,
for a frequency $f\simeq 1~kHz$, the thermal length $l_{ts}$ in the
substrate at low temperature gets of the order of $10~cm$, to be compared
with a room temperature value $l_{ts}(300~K)\simeq 10^{-2}~cm$ (see Table 
\ref{Table_1}). Then at low temperature we rather have 
\begin{equation}
l_{ts}\gtrsim r_{0},  \label{Eq_lts}
\end{equation}
at all frequencies below some $1~kHz$, both for mirrors of
gravitational-wave interferometers (for which $r_{0}\simeq 1.5~cm$), and for
optomechanical sensors (for which $r_{0}\leq 3~10^{-2}~cm$). This value for $%
l_{ts}$ stays constant in the whole region $T\leq 10~K$, as crystalline
materials follow Debye $T^{3}$ laws for $\alpha (T)$, $C(T)$ and $\kappa (T)$%
, and thus their ratios are all independent of $T.$

\begin{table}[th]
\renewcommand{\arraystretch}{1.2} 
\begin{tabular}{c|c|c|c}
Fused silica & $300~K$ & $10~K$ & $1~K$ \\ \hline
$\alpha $ ($K^{-1}$) & $5.5~10^{-7}$ & $-2.6~10^{-7}$ & $-2.6~10^{-10}$ \\ 
$\kappa $ ($W/m.K$) & $1.4$ & $0.1$ & $2~10^{-2}$ \\ 
$C$ ($J/Kg.K$) & $6.7~10^{2}$ & $3$ & $3~10^{-3}$ \\ 
$\lambda$ ($m$) & $8~10^{-10}$ & $8~10^{-8}$ & $9~10^{-6}$ \\ \hline
$\alpha /\kappa $ ($m/W$) & $3.9~10^{-7}$ & $2.6~10^{-6}$ & $1.3~10^{-8}$ \\ 
$l_{t}$ ($m$) & $3~10^{-5}$ & $1.2~10^{-4}$ & $1.7~10^{-3}$%
\end{tabular}
\renewcommand{\arraystretch}{1} \vspace{2mm} \renewcommand{%
\arraystretch}{1.2} 
\begin{tabular}{c|c|c|c}
Sapphire & $300~K$ & $10~K$ & $1~K$ \\ \hline
$\alpha $ ($K^{-1}$) & $5~10^{-6}$ & $5.8~10^{-10}$ & $5.8~10^{-13}$ \\ 
$\kappa $ ($W/m.K$) & $40$ & $4.3~10^{3}$ & $4.3$ \\ 
$C$ ($J/Kg.K$) & $7.9~10^{2}$ & $8.9~10^{-2}$ & $8.9~10^{-5}$ \\ 
$\lambda$ ($m$) & $5~10^{-9}$ & \multicolumn{2}{c}{$2.2~10^{-3}$} \\ \hline
$\alpha /\kappa $ ($m/W$) & $1.2~10^{-7}$ & \multicolumn{2}{c}{$1.4~10^{-13}$%
} \\ 
$l_{t}$ ($m$) & $1.1~10^{-4}$ & \multicolumn{2}{c}{$0.11$}
\end{tabular}
\renewcommand{\arraystretch}{1} \vspace{2mm}
\caption{Thermal properties of fused silica (top) and sapphire (bottom) at
different temperatures. The thermal expansion coefficient $\protect\alpha $,
thermal conductivity $\protect\kappa $, thermal capacity $C$ and phonon mean
free path $\protect\lambda $ are derived from\protect\cite{Braginsky} at
room temperature, and from\protect\cite{Berman,Zeller,Liu2} at low
temperatures. The thermal length $l_{t}$ at $1~kHz$ is obtained from Eq. (%
\ref{Eq_lt}).}
\label{Table_1}
\end{table}

High reflection coatings are typically 40 layers one quarter wavelength
thick of alternating amorphous materials as TiO$_{2}$ and SiO$_{2}$, with a
total thickness $z_{c}\simeq 10^{-3}~cm$ for Nd-Yag laser light. For such a
coating, the breakdown temperature for Eq. (\ref{Eq_ltCond}) is different
for mirrors of large-scale interferometers and for mirrors of high-finesse
cavities, because of the difference in $r_{0}$. For LIGOII mirrors for
instance, taking SiO$_{2}$ as the reference material for the coating, the
thermal length $l_{tc}$ in the coating is of the order of $r_{0}$ only at
very low temperature, $T<1~K$. Amorphous silica films would have $%
l_{tc}>10^{-2}~cm$ for $T<10~K$, so that for high-finesse cavities we have $%
l_{tc}\gtrsim r_{0}$ in the whole low temperature region.

Despite this difference, there are two features of relevance, which affect
similarly the thermal behaviour of the coating-substrate composite in both
types of mirrors. In both cases we have that, at all low temperatures, the
thermal length $l_{tc}$ stays longer than the coating thickness, $%
l_{tc}>z_{c}$, and that the mean free path $\lambda _{s}$ of the phonons in
the substrate is itself long, at least a fraction of $cm$\cite{Berman}. In a
coating of a SiO$_{2}$ film even the phonon mean free path $\lambda _{c}$
will be larger than $10^{-3}~cm$, and thus $\lambda _{c}>z_{c}$, for $%
T\lesssim 1~K$\cite{Vu}.

Let us then consider how the thermal regime changes at sufficiently low
temperatures ($T\leq 10~K$). The heat delivered by the absorbed photons in
the volume $r_{0}^{2}z_{c}$ of the coating crosses to the substrate in a
time smaller than $1/f$, as $l_{tc}>z_{c}$. From the substrate, as $\lambda
_{s}\gtrsim r_{0}$, the thermal phonons thereby created will reenter the
coating, heating it up in even shorter time over distances $\lambda _{s}$.
This happens because the acoustic mismatch between coating and substrate is
small\cite{Little}, when densities and sound velocities are quite close. The
substrate thus acts as a thermal short for the coating in the plane of the
mirror end face: the coating and the substrate will be practically
isothermal over distances of the order of the phonon mean free path $\lambda
_{s}$ in the substrate. Then the coating will contribute to the
thermoelastic fluctuations with its thermal expansion coefficient $\alpha
_{c}$, but following the thermal fluctuations of spectral density $S_{T}$ of
the substrate. On the other hand, at the frequency $f$, the volume of
substrate involved in the fluctuating heating will be of the order $V\simeq
l_{ts}^{3}$, where the thermal length is that in the substrate. So,
including both the coating and the substrate, we write now for the
displacement spectral density $S_{z}\left[ f\right] $: 
\begin{equation}
S_{z}\left[ f\right] \simeq \left( \left( \alpha _{c}z_{c}\right)
^{2}+\left( \alpha _{s}l_{ts}\right) ^{2}\right) S_{T}\left[ f\right] .
\end{equation}
According to table \ref{Table_1}, $\alpha _{c}z_{c}$ is at least one order
of magnitude smaller than $\alpha _{s}l_{ts}$ at low temperature ($%
z_{c}\simeq 10^{-5}~m$ and $l_{ts}\simeq 0.1~m$). We can then neglect the
expansion of the coating over its thickness $z_{c}$ and we find that the
effect is dominated by the substrate properties, 
\begin{equation}
S_{z}\left[ f\right] \simeq \left( \frac{\alpha _{s}}{\rho
_{s}C_{s}l_{ts}^{2}}\right) ^{2}\frac{S_{abs}}{f^{2}}.  \label{Eq_Sxxp2}
\end{equation}
This is the relevant result of our discussion of thermal behaviour of the
coating-substrate composite at low temperature, in that now the temperature
fluctuations involve comparatively large substrate volumes, instead of the
comparatively small coating volume, where the actual absorption of photons
occurs. Notice that, would not this be the case, one would have of course
very large effects just concentrated in the volume, the external surface of
which is that where displacements are going to be measured at SQL
sensitivities.

When we substitute in Eq. (\ref{Eq_Sxxp2}) the expression (\ref{Eq_lt}) for
the thermal length, we see a dramatic change of regime, 
\begin{equation}
S_{z}\left[ f\right] \simeq \left( \frac{\alpha _{s}}{\kappa _{s}}\right)
^{2}S_{abs},  \label{Eq_Sxxp3}
\end{equation}
where now the (substrate) thermal conductivity $\kappa _{s}$ appears,
instead of the thermal capacity, and the frequency dependence has
disappeared.\ In fact the system behaves as in the low frequency region of a
low pass filter, while, under BGV conditions, it was rather in the high
frequency region.

We develop in the following sections a rigorous calculation of the effects
in the low temperature regime, which gives in clear details the features
grossly anticipated above and which can be directly applied to mirrors of
interest for optomechanical devices, when the above thermal behaviour of the
coating-substrate system is realized.

\section{Thermodynamic noise}

\label{Thermodynamic}In this section we determine the thermodynamic noise
without any assumption on the ratio $l_{t}/r_{0}$ between the thermal
diffusion length in the substrate and the beam spot size. Our analysis is an
extension of the procedure developed by Liu and Thorne\cite{Liu}, but it is
valid even when the adiabatic approximation is not satisfied. According to
Eq.\ (\ref{Eq_lt}), the condition $l_{t}<r_{0}$ can actually be written as a
condition over the frequency, $f>\kappa /\rho Cr_{0}^{2}$. Our treatment is
thus also valid for an angular frequency $\omega =2\pi f$ smaller than the
adiabatic limit $\omega _{c}$ defined as 
\begin{equation}
\omega _{c}=\frac{\kappa }{\rho Cr_{0}^{2}}.  \label{Eq_DefOmegac}
\end{equation}

As shown in the previous section we neglect any effects of the coating,
taking only into account the thermal properties of the substrate of the
mirror. We also neglect any finite-size effects since we have shown that the
volume of substrate involved in the fluctuating heating is usually smaller
than the size of the mirror, even at low temperature. We thus approximate
the mirror as a half space, the coated plane face corresponding to the plane 
$z=0$ in cylindrical coordinates.

The analysis is based on a general formulation of the
fluctuation-dissipation theorem, used by Levin\cite{Levin} to compute the
usual thermal noise (brownian motion) of the mirrors in a gravitational-wave
interferometer. We know that in an interferometer or in a high finesse
Fabry-Perot cavity, the light is sensitive to the normal displacement $%
u_{z}\left( z=0,{\bf r},t\right) $ of the coated plane face of the mirror,
spatially averaged over the beam profile.\ This averaged displacement $%
\widehat{u}$ is defined as 
\begin{equation}
\widehat{u}(t)=\int d^{2}r~u_{z}\left( z=0,{\bf r},t\right) \frac{%
e^{-r^{2}/r_{0}^{2}}}{\pi r_{0}^{2}}.  \label{mean displacement}
\end{equation}
To compute the spectral density $S_{\widehat{u}}\left[ \omega \right] $ of
the displacement $\widehat{u}$ at a given angular frequency $\omega $, we
determine the mechanical response of the mirror to a sinusoidally
oscillating pressure.\ More precisely, we examine the effect of a pressure $%
P\left( {\bf r},t\right) $ applied at every point ${\bf r}$ of the coated
face of the mirror with the same spatial profile as the optical beam, 
\begin{equation}
P\left( {\bf r},t\right) =\frac{F_{0}}{\pi r_{0}^{2}}e^{-r^{2}/r_{0}^{2}}%
\cos \left( \omega t\right) ,  \label{Pressure}
\end{equation}
where $F_{0}$ is a constant force amplitude. We can compute the energy $%
W_{diss}$ dissipated by the mirror in response to this force, averaged over
a period $2\pi /\omega $ of the pressure oscillations.\ The
fluctuation-dissipation theorem then states that the spectral density of the
displacement noise is given by 
\begin{equation}
S_{\widehat{u}}\left[ \omega \right] =\frac{8k_{B}T}{\omega ^{2}}\frac{%
W_{diss}}{F_{0}^{2}},  \label{FluctuationDissipation}
\end{equation}
where $k_{B}$ is the Boltzman's constant. This approach has been used by
Levin to compute the brownian noise\cite{Levin}.\ We are interested here in
the thermodynamic noise so that $W_{diss}$ corresponds to the energy
dissipated by thermoelastic heat flow.

The rate of thermoelastic dissipation is given by the following expression
(first term of Eq. (35.1) of Ref.\cite{Landau}): 
\begin{equation}
W_{diss}=\left\langle T\frac{dS}{dt}\right\rangle =\left\langle \int d^{3}r~%
\frac{\kappa }{T}\left( {\bf \nabla }\delta T\right) ^{2}\right\rangle ,
\label{Wdissipation}
\end{equation}
where the integral is on the entire volume of the mirror and the brackets $%
\left\langle ...\right\rangle $ stand for an average over the oscillation
period $2\pi /\omega $. $\delta T$ is the temperature perturbation around
the unperturbed value $T$, induced by the oscillating pressure. $W_{diss}$
is then related to the time derivative $dS/dt$ of the mirror's entropy,
which depends on the temperature gradient ${\bf \nabla }\delta T$.

To calculate the rate of energy dissipation $W_{diss}$, it is necessary to
solve a system of two coupled equations, the first one for the displacement $%
{\bf u}\left( {\bf r},t\right) $ at every point ${\bf r}$ inside the
substrate, and the second one for the temperature perturbation $\delta
T\left( {\bf r},t\right) $. As the time required for sound to travel across
the mirror is usually smaller than the oscillation period $2\pi /\omega $,
we can use a quasistatic approximation and deduce the displacement ${\bf u}$
from the equation of static stress balance\cite{Landau}, 
\begin{equation}
{\bf \nabla }\left( {\bf \nabla }.{\bf u}\right) +\left( 1-2\sigma \right)
\nabla ^{2}{\bf u}=-2\alpha \left( 1+\sigma \right) {\bf \nabla }\delta T,
\label{staticstressbalance}
\end{equation}
where $\sigma $ is the Poisson ratio of the substrate ($\alpha $ is the
linear thermal expansion coefficient). The temperature perturbation $\delta
T $ evolves according to the thermal conductivity equation\cite{Landau}, 
\begin{equation}
\frac{\partial \left( \delta T\right) }{\partial t}-a^{2}\Delta \left(
\delta T\right) =\frac{-\alpha ET}{\rho C\left( 1-2\sigma \right) }\frac{%
\partial \left( {\bf \nabla }.{\bf u}\right) }{\partial t},
\label{Eq.temperature}
\end{equation}
where $a^{2}=\kappa /\rho C$ and $E$ is the Young modulus of the substrate ($%
\rho $ is the density, $C$ is the specific thermal capacity).

The solutions of Eqs. (\ref{staticstressbalance}) and (\ref{Eq.temperature})
must also fulfill boundary conditions. If we approximate the mirror as a
half space, the temperature perturbation $\delta T$ and the stress tensor $%
\sigma _{ij}$ must satisfy the following boundary conditions on the coated
plane face of the mirror, 
\begin{mathletters}
\begin{eqnarray}
\sigma _{zz}\left( z=0,{\bf r},t\right)  &=&P\left( {\bf r},t\right) , \\
\sigma _{zx}\left( z=0,{\bf r},t\right)  &=&\sigma _{zy}\left( z=0,{\bf r}%
,t\right) =0, \\
\frac{\partial \left( \delta T\right) }{\partial z}\left( z=0,{\bf r}%
,t\right)  &=&0.
\end{eqnarray}
The stress tensor $\sigma _{ij}$ is defined in presence of changes of
temperature as (see Eq. (6.2) of \cite{Landau}) 
\end{mathletters}
\begin{eqnarray}
\sigma _{ij} &=&-\frac{E}{\left( 1-2\sigma \right) }\alpha \delta T\delta
_{ij}+  \nonumber \\
&&+\frac{E}{\left( 1+\sigma \right) }\left[ u_{ij}+\frac{\sigma }{\left(
1-2\sigma \right) }\delta _{ij}\sum_{k}u_{kk}\right] ,  \label{StressTensor}
\end{eqnarray}
where the strain tensor $u_{ij}$ is equal to $\frac{1}{2}\left( \frac{%
\partial u_{i}}{\partial x_{j}}+\frac{\partial u_{j}}{\partial x_{i}}\right) 
$.

We solve perturbatively this system of equations at the first order in $%
\alpha $. We first solve the static stress-balance equation at the zeroth
order in $\alpha $, neglecting the temperature term in the right part of Eq.
(\ref{staticstressbalance}) and in the expression (\ref{StressTensor}) of
the stress tensor. The solution ${\bf u}^{\left( 0\right) }$ of this
equation is well known (paragraph 8 of \cite{Landau}). We then solve the
thermal conductivity equation (\ref{Eq.temperature}) using as a source term
the solution ${\bf u}^{\left( 0\right) }$ and we obtain the temperature
perturbation $\delta T^{\left( 1\right) }$ in the first order in $\alpha $.
The calculation of ${\bf u}^{\left( 0\right) }$, $\delta T^{\left( 1\right)
} $ and finally $W_{diss}$ is done in Appendix A. Using the results of this
appendix, we show that $S_{\widehat{u}}\left[ \omega \right] $ is equal to

\begin{equation}
S_{\widehat{u}}\left[ \omega \right] =32\alpha ^{2}\left( 1+\sigma \right)
^{2}\frac{k_{B}T^{2}}{\rho C}I,  \label{SuOmega1}
\end{equation}
where the integral $I$ is given by 
\begin{equation}
I=\frac{a^{2}}{(2\pi )^{3}}\int dk_{x}dk_{y}dk_{z}\frac{k_{\perp
}^{2}e^{-k_{\perp }^{2}r_{0}^{2}/2}}{k^{2}\left( a^{4}k^{4}+\omega
^{2}\right) },  \label{I}
\end{equation}
with $k_{\perp }^{2}=k_{x}^{2}+k_{y}^{2}$ and $k^{2}=k_{\perp
}^{2}+k_{z}^{2}.$

We can express $S_{\widehat{u}}\left[ \omega \right] $ as a function of an
integral $J\left[ \Omega \right] $ which depends only on a dimensionless
variable $\Omega $ equal to $\omega /\omega _{c}$, where $\omega
_{c}=a^{2}/r_{0}^{2}$ corresponds to the adiabatic limit (see Eq.\ \ref
{Eq_DefOmegac}). We get 
\begin{equation}
S_{\widehat{u}}\left[ \omega \right] =\frac{8}{\sqrt{2\pi }}\alpha
^{2}\left( 1+\sigma \right) ^{2}\frac{k_{B}T^{2}r_{0}}{\rho Ca^{2}}J\left[
\Omega \right] ,  \label{SuOmega2}
\end{equation}
where $J[\Omega ]$ is derived from the integral $I$ by the transformation of
variables $u\equiv k_{\perp }r_{0}$ and $v\equiv k_{z}r_{0}$,

\begin{equation}
J\left[ \Omega \right] =\sqrt{\frac{2}{\pi }}\int_{0}^{\infty
}du\int_{-\infty }^{\infty }dv\frac{u^{3}e^{-u^{2}/2}}{\left(
u^{2}+v^{2}\right) \left( \left( u^{2}+v^{2}\right) ^{2}+\Omega ^{2}\right) }%
.  \label{JOmega}
\end{equation}

When $\omega \gg \omega _{c}$ (i.e. $\Omega \gg 1$), we can neglect $%
u^{2}+v^{2}$ with respect to $\Omega $ in the denominator of the integral. $J%
\left[ \Omega \right] $ can then be calculated analytically and we obtain 
\begin{equation}
J\left[ \Omega \gg 1\right] =1/\Omega ^{2}.
\end{equation}
Using this result and the definition of $\Omega $, we finally show that $S_{%
\widehat{u}}\left[ \omega \right] $ is equal to 
\begin{equation}
S_{\widehat{u}}\left[ \omega \gg \omega _{c}\right] =\frac{8}{\sqrt{2\pi }}%
\alpha ^{2}\left( 1+\sigma \right) ^{2}\frac{k_{B}T^{2}}{\rho C}\frac{a^{2}}{%
\omega ^{2}r_{0}^{3}}.
\end{equation}
This formula is identical to the expression (18) of Ref. \cite{Liu} and to
the expression (12) of BGV\cite{Braginsky}.

\begin{figure}[ht]
\centerline{\psfig{figure=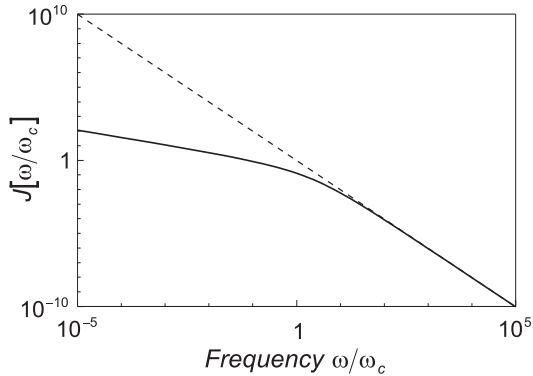,width=7cm}}
\vspace{2mm}
\caption{Frequency dependence of the thermodynamic noise. The
frequency $\omega $ is normalized to the adiabatic limit $\omega _{c}$. The
dashed curve corresponds to the adiabatic approximation.}
\label{Fig_Thermodynamic}
\end{figure}

For small values of $\Omega $, the integral $J\left[ \Omega \right] $ can be
computed numerically. The result is shown in figure \ref{Fig_Thermodynamic}
where we have plotted $J\left[ \Omega \right] $ as a function of $\Omega $
in logarithmic scale, for $\Omega $ between $10^{-5}$ and $10^{5}$. This
figure shows that $\Omega =1$ is a cut-off frequency. For $\Omega >1$ the
curve has a slope equal to $-2$, whereas for $\Omega <1$ the curve has a
smaller slope of the order of $-1/2$. In this low frequency range, the noise
is smaller than the one which would be obtained using the adiabatic
approximation (dashed curve in figure \ref{Fig_Thermodynamic}).

\section{Photothermal noise}

\label{PhotoThermal}We now briefly examine the case of the photothermal
noise which exhibits somewhat a similar frequency behaviour as the
thermodynamic noise. We use the same method as BGV\cite{Braginsky} to
calculate the spectral density $S_{\widehat{u}}\left[ \omega \right] $ due
to this noise but we do not make any adiabatic approximation so that the
calculation is valid also for frequencies smaller than the adiabatic limit $%
\omega _{c}$. We then obtain: 
\begin{equation}
S_{\widehat{u}}\left[ \omega \right] =\frac{2}{\pi ^{2}}\alpha ^{2}\left(
1+\sigma \right) ^{2}\frac{\hbar \omega _{0}W_{abs}}{\left( \rho
Ca^{2}\right) ^{2}}K\left[ \Omega \right] ,  \label{Su(w)}
\end{equation}
where the integral $K\left[ \Omega \right] $ is equal to 
\begin{equation}
K\left[ \Omega \right] =\left| \frac{1}{\pi }\int_{0}^{\infty
}du\int_{-\infty }^{\infty }dv\frac{u^{2}e^{-u^{2}/2}}{\left(
u^{2}+v^{2}\right) \left( u^{2}+v^{2}+i\Omega \right) }\right| ^{2}.
\end{equation}
When $\omega \gg \omega _{c}$ the adiabatic approximation is valid and the
result of BGV\ should be recovered. Indeed, when $\Omega \gg 1,$we can
neglect $u^{2}+v^{2}$ with respect to $\Omega $ and calculate analytically $K%
\left[ \Omega \right] $ which turns out to be equal to $1/\Omega ^{2}$.
Using the definition of $\Omega $, we finally show that $S_{\widehat{u}}%
\left[ \omega \right] $ is equal to 
\begin{equation}
S_{\widehat{u}}\left[ \omega \gg \omega _{c}\right] =\frac{2}{\pi ^{2}}%
\alpha ^{2}\left( 1+\sigma \right) ^{2}\frac{\hbar \omega _{0}W_{abs}}{%
\left( \rho Cr_{0}^{2}\omega \right) ^{2}}.
\end{equation}
This formula is identical to Eq. (8) of BGV\cite{Braginsky}.

\begin{figure}[ht]
\centerline{\psfig{figure=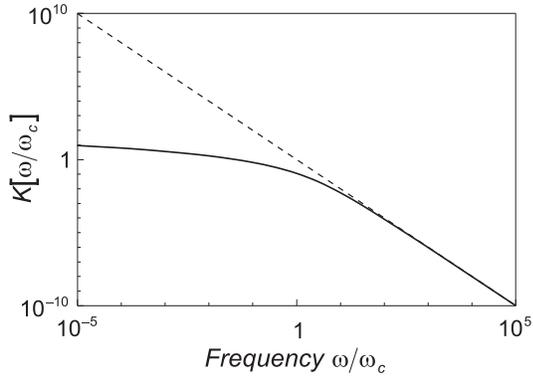,width=7cm}}
\vspace{2mm}
\caption{Frequency dependence of the photothermal noise. The
frequency $\omega $ is normalized to the adiabatic limit $\omega _{c}$. The
dashed curve corresponds to the adiabatic approximation.}
\label{Fig_Photothermal}
\end{figure}

For low values of $\Omega $, $K\left[ \Omega \right] $ can be calculated
numerically. The result is shown in figure \ref{Fig_Photothermal}. As in the
case of the thermodynamic effect (figure \ref{Fig_Thermodynamic}), $\Omega
=1 $ is a cut-off frequency: for $\Omega >1$ the function has a slope equal
to $-2$, whereas for $\Omega <1$ the function has a much smaller slope and
is almost constant.

This result is in perfect agreement with the simple derivation made in
section \ref{HeatPropag}.\ The spectral density (\ref{Su(w)}) can actually
be written as 
\begin{equation}
S_{\widehat{u}}\left[ \omega \right] =\frac{2}{\pi ^{2}}\left( 1+\sigma
\right) ^{2}\left( \frac{\alpha }{\kappa }\right) ^{2}S_{abs}K\left[ \Omega %
\right] ,  \label{Su(w)2}
\end{equation}
which is similar to Eq. (\ref{Eq_Sxxp3}) at low frequency where $K\left[
\Omega \right] \simeq 1$, apart from a term with the Poisson ratio. The
frequency dependence of the photothermal noise then corresponds to a
low-pass filter, with a cut-off frequency equal to the adiabatic limit $%
\omega _{c}$.\ At low frequency, that is when the thermal diffusion length $%
l_{t}$ in the substrate becomes larger than the beam spot radius $r_{0}$,
one has a dramatic change of regime and the photothermal noise is much
smaller than the one which would be obtained within the adiabatic
approximation (dashed curve in figure \ref{Fig_Photothermal}).

Another important point for the realization of optomechanical sensors
working at the quantum level is to compare the photothermal noise to the
displacements induced by the quantum fluctuations of radiation pressure of
light. Quantum effects can be made larger than the usual thermal noise by
decreasing the temperature and by increasing the light power.\ This would
not be convenient for photothermal noise since the effect is proportional to
the light power. BGV\ results show that for sapphire at room temperature the
photothermal noise is of the same order as the standard quantum limit in an
interferometer.\ In a high-finesse cavity, hopefully, these two effects are
related to quite different photon statistics.\ As a matter of fact, the
displacement $\widehat{u}$ induced by the radiation pressure of the
intracavity field is related to the intracavity photon flux $N$, 
\begin{equation}
\widehat{u}\left[ \omega \right] =\chi _{eff}\left[ \omega \right] P_{rad}%
\left[ \omega \right] =2\hbar k\chi _{eff}\left[ \omega \right] N\left[
\omega \right] ,  \label{Suscep.Prad}
\end{equation}
where $2\hbar k$ is the momentum exchange during a photon reflection ($k$ is
the wavevector of light), and $\chi _{eff}$ is an effective susceptibility
describing the mechanical response of the mirror to the radiation pressure $%
P_{rad}$\cite{Heidmann97}.\ The noise spectrum $S_{\widehat{u}}\left[ \omega %
\right] $ induced by radiation pressure is thus proportional to the spectral
power noise $S_{cav}$ of the intracavity light, which for a resonant cavity
is\cite{Heidmann97}: 
\begin{equation}
S_{cav}\left[ \omega \right] =\frac{2{\cal F}/\pi }{1+\left( \omega /\omega
_{cav}\right) ^{2}}\hbar \omega _{0}W_{cav},  \label{Eq_Scav}
\end{equation}
where $\omega _{cav}$ is the cavity bandwidth, ${\cal F}$ is the cavity
finesse and $W_{cav}=\hbar \omega _{0}\overline{N}$ is the average
intracavity light power. At low frequency ($\omega \lesssim \omega _{cav}$)
the intracavity photon flux corresponds to a super-poissonian statistics,
the noise power being larger than the poissonian spectral density by a
factor $2{\cal F}/\pi $\cite{Teich}.

On the other hand, the absorbed photons always corresponds to a poissonian
statistics, even if it is not the case for the intracavity photons. The
spectral power noise $S_{abs}$ of the absorbed light is given by 
\begin{equation}
S_{abs}=\hbar \omega _{0}W_{abs}=A\hbar \omega _{0}W_{cav},  \label{Eq_Sabs}
\end{equation}
where $A$ is the absorption coefficient of the mirror (the average flux of
absorbed photons is $\overline{n}=A\overline{N}$).\ This effect cannot be
understood within the framework of a corpuscular model in which the photon
absorption is described as a poissonian process: due to the super-poissonian
statistics of the intracavity photons, one would find a super-poissonian
statistics for the absorbed photons. One has to take into account the
interferences between the intracavity field and the vacuum fluctuations
associated with the mirror losses. This can be done by using a simple model
where the absorption is described as a small transmission of the mirror and
where\ the absorbed photons are identified to the photons transmitted by the
mirror. One thus has a high-finesse cavity with two input-output ports and
it is well known that the photon statistics of the light either reflected or
transmitted by such a cavity are always poissonian, for coherent or vacuum
incoming beams\cite{Reynaud}.

Eqs. (\ref{Eq_Scav}) and (\ref{Eq_Sabs}) clearly show that both the
radiation pressure effect and the photothermal noise are proportional to the
intracavity light power $W_{cav}$; however the displacements induced by
radiation pressure have an extra dependence on the cavity finesse ${\cal F}$%
. The photothermal noise can thus become negligible as compared to quantum
effects for a high-finesse cavity.

To perform a quantitative comparison between the two effects, we calculate
the susceptibility $\chi _{eff}$ defined by Eq. (\ref{Suscep.Prad}). We
determine here the mechanical response associated with the internal degrees
of freedom of the mirror, which are of interest for displacement sensors. We
thus ignore the radiation pressure effects associated with the global motion
of suspended mirrors, which are the dominant contribution to SQL at low
frequency in gravitational-wave interferometers. We calculate the average
displacement $\widehat{u}$ induced by the radiation pressure $P_{rad}$,
assuming the mirror is a half space ($z\geq 0$). The normal displacement $%
u_{z}(z=0,{\bf r},t)$ of the coated face of the mirror can be deduced from
the results of paragraph 8 of \cite{Landau}, 
\begin{equation}
u_{z}(z=0,{\bf r},t)=\frac{2\hbar kN\left( t\right) }{E\pi ^{2}r_{0}^{2}}%
\left( 1-\sigma ^{2}\right) \int d^{2}r^{\prime }\frac{e^{-r^{\prime
2}/r_{0}^{2}}}{\left| {\bf r}-{\bf r}^{\prime }\right| }.
\end{equation}
Using the definition (\ref{mean displacement}) of $\widehat{u}$, we obtain 
\begin{equation}
\widehat{u}\left( t\right) =\frac{2\hbar kN\left( t\right) }{E\pi
^{3}r_{0}^{4}}\left( 1-\sigma ^{2}\right) \int d^{2}rd^{2}r^{\prime }\frac{%
e^{-\left( r^{2}+r^{\prime 2}\right) /r_{0}^{2}}}{\left| {\bf r}-{\bf r}%
^{\prime }\right| }.
\end{equation}
The integral can easily be calculated by using a new set of variables ${\bf u%
}={\bf r}-{\bf r}^{\prime }$ and ${\bf v}={\bf r}+{\bf r}^{\prime }$. We
finally get 
\begin{equation}
\chi _{eff}\left[ \omega \right] =\frac{1-\sigma ^{2}}{\sqrt{2\pi }Er_{0}},
\end{equation}
and the noise spectrum $S_{\widehat{u}}\left[ \omega \right] $ induced by
radiation pressure fluctuations is equal to 
\begin{equation}
S_{\widehat{u}}\left[ \omega \right] =\left( \frac{2\left( 1-\sigma
^{2}\right) }{\sqrt{2\pi }Ecr_{0}}\right) ^{2}S_{cav}\left[ \omega \right] ,
\label{Su(w)prad}
\end{equation}
where $c$ is the speed of light.

For all the displacement sensors considered in this paper the characteristic
angular frequency $\omega $ is smaller than the cavity bandwidth $\omega
_{cav}$. The noise spectrum $S_{\widehat{u}}\left[ \omega \right] $ is
consequently independent of $\omega $ and equal to 
\begin{equation}
S_{\widehat{u}}\left[ \omega \ll \omega _{cav}\right] =\left( \frac{2\left(
1-\sigma ^{2}\right) }{\sqrt{2\pi }Ecr_{0}}\right) ^{2}\frac{2{\cal F}}{\pi }%
\hbar \omega _{0}W_{cav}.  \label{Su(w)pradLF}
\end{equation}
This expression shows that the radiation pressure effect depends on the
mechanical characteristics of the substrate ($E$ and $\sigma $) whereas the
photothermal noise (Eq. \ref{Su(w)2}) depends on the thermodynamic
characteristics of the substrate via the ratio $\alpha /\kappa $. At low
temperature, $K\left[ \Omega \right] $ is of the order of $1$ and the ratio $%
\alpha /\kappa $ is constant and equal to $1.4~10^{-13}~m/W$ for sapphire
(see Table \ref{Table_1}). $E$ is equal to $4~10^{11}~J/m^{3}$ and $\sigma $
is equal to $0.25$ so that the ratio between the photothermal and radiation
pressure noises is of the order of 
\begin{equation}
S_{\widehat{u}}^{pt}/S_{\widehat{u}}^{rad}\simeq 2.5~10^{14}\frac{Ar_{0}^{2}%
}{{\cal F}}.
\end{equation}
For a $1~ppm$ absorption rate ($A=10^{-6}$), a beam spot size $r_{0}$ of $%
10^{-4}~m$, and a cavity finesse ${\cal F}$ of $10^{5}$, the photothermal
noise is more than 4 orders of magnitude smaller than the radiation pressure
effects of internal degrees of freedom of the mirror. The photothermal noise
is thus negligible as compared to quantum effects in optomechanical sensors.

Note that this is not the case in gravitational-wave interferometers where $%
r_{0}\simeq 10^{-2}~m$ and ${\cal F}\simeq 100$. The photothermal noise is
then 2 orders of magnitude larger than the quantum noise of internal
motion.\ However, the interferometer is not expected to be sensitive to this
quantum noise since for suspended mirrors it is overwhelmed by the quantum
noise associated with external pendulum motion.

\section{Discussion and conclusion}

\label{Discussion}We have shown that both thermoelastic and photothermal
noises have a frequency dependence which looks like a low-pass filter: below
a cut-off frequency $\omega _{c}$, these noises are much smaller than the
noise which would be obtained according to the $1/\omega ^{2}$ dependence at
high-frequency.

First lines in tables \ref{Table_2} give the values of the cut-off frequency%
{\small \ }$\omega _{c}/2\pi $ for fused silica and sapphire, and for a beam
spot size $r_{0}$ of $1$~$cm$ (first table) and $10^{-2}~cm$ (second table).
The results show that $\omega _{c}$ is increased when the temperature
decreases (3 orders of magnitude for fused silica and 6 orders of magnitude
for sapphire when the temperature is reduced from $300~K$ to $1~K$). If we
consider a typical frequency $\omega /2\pi $ of $100~Hz$, the adiabatic
approximation is never valid for sapphire at low temperature, whereas it is
valid for fused silica only for large beam spot size.

\begin{table}[th]
\renewcommand{\arraystretch}{1.2} 
\begin{tabular}{c|c|c|c|c}
$r_{0}=10^{-2}~m$ & \multicolumn{2}{c|}{Fused silica} & \multicolumn{2}{c}{
Sapphire} \\ 
& $300~K$ & $1~K$ & $300~K$ & $1~K$ \\ \hline
$\omega _{c}/2\pi $ ($Hz$) & $1.5~10^{-3}$ & $4.8$ & $2~10^{-2}$ & $%
1.9~10^{4}$ \\ 
$\Omega ^{2}J\left[ \Omega \right] $ & $1$ & $0.74$ & $0.98$ & $1.3~10^{-4}$
\\ 
$S_{\widehat{u}}$ ($m^{2}/Hz$) & $2.7~10^{-42}$ & $3.4~10^{-45}$ & $%
1.5~10^{-39}$ & $2.6~10^{-49}$ \\ 
$S_{\widehat{u}}$ (adiabatic) & $2.7~10^{-42}$ & $4.6~10^{-45}$ & $%
1.5~10^{-39}$ & $2~10^{-45}$%
\end{tabular}
\renewcommand{\arraystretch}{1} \vspace{2mm} \renewcommand{%
\arraystretch}{1.2} 
\begin{tabular}{c|c|c|c|c}
$r_{0}=10^{-4}~m$ & \multicolumn{2}{c|}{Fused silica} & \multicolumn{2}{c}{
Sapphire} \\ 
& $300~K$ & $1~K$ & $300~K$ & $1~K$ \\ \hline
$\omega _{c}/2\pi $ ($Hz$) & $15$ & $4.8~10^{4}$ & $2~10^{2}$ & $1.9~10^{8}$
\\ 
$\Omega ^{2}J\left[ \Omega \right] $ & $0.51$ & $3.5~10^{-5}$ & $6.4~10^{-2}$
& $1.4~10^{-10}$ \\ 
$S_{\widehat{u}}$ ($m^{2}/Hz$) & $1.4~10^{-36}$ & $1.6~10^{-43}$ & $%
1~10^{-34}$ & $2.8~10^{-49}$ \\ 
$S_{\widehat{u}}$ (adiabatic) & $2.7~10^{-36}$ & $4.6~10^{-39}$ & $%
1.6~10^{-33}$ & $2~10^{-39}$%
\end{tabular}
\renewcommand{\arraystretch}{1} \vspace{2mm}
\caption{Results for fused silica and sapphire at different temperatures,
for a frequency $\protect\omega /2\protect\pi =100~Hz$ and for a beam spot
size $r_{0}=1~cm$ (top) and $10^{-2}~cm$ (bottom). The thermodynamic noise $%
S_{\widehat{u}}$ (3$^{rd}$ lines) is reduced as compared to its value
obtained within the adiabatic approximation (4$^{th}$ lines) by a factor $%
\Omega ^{2}J\left[ \Omega \right] $ (2$^{nd}$ lines).}
\label{Table_2}
\end{table}

We first focus on the thermodynamic noise whose values calculated from Eq. (%
\ref{SuOmega2}) are shown in the third lines of tables \ref{Table_2}. The
noise is smaller than the one which would be obtained within the adiabatic
approximation (last lines in tables \ref{Table_2}). The reduction factor,
equal to $1/\Omega ^{2}J\left[ \Omega \right] $, can be as large as $10^{4}$
for sapphire at low temperature with $r_{0}=1~cm$, and as large as $10^{10}$
for $r_{0}=10^{-2}~cm$ (second lines in tables \ref{Table_2}).

We immediately see the impact for gravitational-wave interferometers ($%
r_{0}=1~cm$): for sapphire at low temperature, the thermodynamic noise is
more than 4 order of magnitude smaller than for fused silica, so that the
choice of the material at low temperature would be just the opposite than,
as in BGV, at room temperature. Furthermore, the thermodynamic noise at $%
100~Hz$ would be equal to $2.6~10^{-49}~m^{2}/Hz$ for sapphire at low
temperature, well below the noise at room temperature ($%
2.7~10^{-42}~m^{2}/Hz $ for fused silica).\ It is also well below the SQL
limit due to the external pendulum motion, equal to $3.6~10^{-41}~m^{2}/Hz$
for a mirror mass of $30~kg$\cite{Braginsky}.

Similarly for optomechanical systems with smaller beam spot size ($%
r_{0}\lesssim 10^{-2}~cm$), the thermodynamic noise can be made as small as $%
2.8~10^{-49}~m^{2}/Hz$ by using sapphire at low temperature, to be compared
to a noise larger than $10^{-36}~m^{2}/Hz$ at room temperature both for
sapphire and fused silica. It is worth noticing that this very low value is
partly due to the reduction factor associated with the non adiabaticity
which is of the order of $10^{10}$. This noise can be compared to the SQL\
limit due to the internal motion of the mirror, which is equal to $\hbar
\left| \chi _{eff}\right| \simeq 10^{-42}~m^{2}/Hz$\cite{Jaekel}. At low
temperature, the thermodynamic noise is thus smaller than the SQL\ limit so
that optomechanical sensors as in Refs. \cite{Conti,Cohadon} would be able
to get to the SQL\ limit.

Let us note that the thermodynamic noise for sapphire at $1~K$ is mostly
independent of the beam spot size $r_{0}$: similar values are obtained for
large spot sizes ($2.6~10^{-49}~m^{2}/Hz$ for $r_{0}=1~cm$) and small ones ($%
2.8~10^{-49}~m^{2}/Hz$ for $r_{0}=10^{-2}~cm$). This result is due to the
fact that, in contrast with fused silica, the adiabatic approximation is not
valid for sapphire whatever the beam spot size is, as long as it is smaller
than a few centimeters. The non adiabatic condition $\omega <\omega _{c}$
can actually be written as $r_{0}<\sqrt{\kappa /\rho C\omega }$ (see Eq.\ 
\ref{Eq_DefOmegac}). In this non adiabatic regime, we have shown that $J%
\left[ \Omega \right] $ evolves as $1/\sqrt{\Omega }$ which is proportional
to $\sqrt{\omega _{c}}$ and then to $1/r_{0}$. The thermodynamic noise is
proportional to $r_{0}J\left[ \Omega \right] $ (Eq. \ref{SuOmega2}) and is
then independent of $r_{0}$.

Similar results can be obtained for the photothermal noise. We have shown
that as long as the adiabatic condition is not satisfied, the noise mostly
depends on the ratio $\alpha /\kappa $ (Eqs. \ref{Eq_Sxxp3} or \ref{Su(w)2}%
). In particular it does not depend on the frequency $\omega $ nor on the
beam spot size $r_{0}$. For sapphire at low temperature and for an average
absorbed power $W_{abs}\simeq 1~W$ of Nd-Yag laser light, the photothermal
noise is then of the order of $10^{-45}~m^{2}/Hz$ both for interferometers
and optomechanical sensors, well below the SQL limits of both the external
and internal motions.

Let us finally note that the results obtained above apply in detail to an
actual mirror system when the conditions described in section \ref
{HeatPropag} are fulfilled. In particular the interplay of the various
characteristic lengths (phonon mean free path and thermal lengths at the
frequencies of interest, both in the substrate and in the coating, beam
spot, coating thickness and mirror size) must in the end allow, in some
temperature range, that the thermal properties of the substrate dominate.

This may be not easy to achieve and thus our analysis may correspond to a
somewhat idealized situation. At the lowest temperatures $T\leq 0.5~K$, the
phonon mean free path in the coating gets of the order of its thickness (as
for an amorphous silica coating, see for example \cite{Vu}), while the time
constants for phonon local equilibrium continue to stay smaller than $2\pi
/\omega $ both for the coating and the substrate. At first sight, it may
appear that the equalization of coating versus substrate temperatures will
be even more facilitated. However at some intermediate temperature below $%
10~K$ the phonon mean free path in substrates like sapphire may get so long
to exceed the dimensions of the mirror. In this case a more ad hoc model has
to be considered, in which one specifies the details of the thermal link of
the mirror to the main heat sink. Similar care should be taken if the
thermal length at the lowest frequency of interest is of the order of the
mirror size. Possibly in both cases the effect would be smaller than
predicted in this paper, because the characteristic times for thermal
equilibrium in the mirror volume would get even shorter. In any case, as to
quote just one instance, phonon mean free paths and thermal conductivities
(and thus thermal lengths), have strong dependencies at low temperatures on
the level of impurities, so that each experimental configuration may be a
case per se.

In conclusion our results may be possibly of interest in two respects. First
because they let see promising the use of low temperatures both for
gravitational-wave interferometers and for optomechanical devices. Second
because they may be of help to study, in actual experimental configurations
for SQL conditions, the role of the thermoelastic effects in respect to
choices of substrate materials, working temperatures, cavity finesses,
mirrors losses, beam spot sizes and laser powers.

\section*{Acknowledgements}

After the completion of this work, we learned that results for the
photothermal effect similar to those of section \ref{PhotoThermal} were
obtained with a different method\cite{Vyatchanin}, that the problem of
section \ref{Thermodynamic} has been addressed heuristically with conclusions in part
similar \cite{Rowan}, and that a calculation of thermoelastic effects due to
photon absorption in the bulk of cryogenic crystallin cavities\cite{Liu2}
was also having similar features. We are very grateful to Vladimir Braginsky
and Sergey Vyatchanin for their private communication and to Sheila Rowan
and to Stephan Schiller for drawing attention to their unpublished results.
M.\ Pinard is grateful to INFN Legnaro National Laboratories for
hospitality during a short visit within the E.U. programme "TMR - Access to
LSF" (Contract ERBFMGECT980110).
  
\section*{Appendix A}

In this appendix, we calculate the spectral density $S_{\widehat{u}}\left[
\omega \right] $ (Eq. \ref{SuOmega1}) of the spatially averaged displacement 
$\widehat{u}$ induced by the thermodynamic noise. We approximate the mirror
as an infinite half space ($z\geq 0$). At the zeroth order in the thermal
expansion coefficient $\alpha $, the solution ${\bf u}^{\left( 0\right) }$
of the quasistatic stress balance equation (\ref{staticstressbalance}) is
given by a Green's tensor (Eqs. (8.13) and (8.18) of \cite{Landau}). The
pressure $P\left( {\bf r},t\right) $ applied on the coated surface of the
mirror has only a component along the normal axis $z$, and we obtain the
following expression for the displacement expansion $\Theta ^{\left(
0\right) }={\bf \nabla }.{\bf u}^{\left( 0\right) }$, 
\begin{eqnarray}
\Theta ^{\left( 0\right) }\left( {\bf r},t\right) &=&-\frac{2\left( 1+\sigma
\right) \left( 1-2\sigma \right) }{E}F_{0}\cos \left( \omega t\right) \times
\nonumber \\
&&\times \int \frac{dk_{x}dk_{y}}{\left( 2\pi \right) ^{2}}e^{-%
{\frac14}%
k_{\perp }^{2}r_{_{0}}^{2}-k_{\perp }z+i\left( k_{x}x+k_{y}y\right) },
\end{eqnarray}
with $k_{\perp }=\sqrt{k_{x}^{2}+k_{y}^{2}}$.

To calculate the dissipated energy $W_{diss}$ it is useful to analytically
extend the pressure-induced expansion $\Theta ^{\left( 0\right) }$ for
negative values of $z$ in such a way that it is an even function of $z$. Its
spatial Fourier transform $\Theta ^{\left( 0\right) }\left[ {\bf k},t\right] 
$ is then equal to 
\begin{equation}
\Theta ^{\left( 0\right) }\left[ {\bf k},t\right] =-\frac{4\left( 1+\sigma
\right) \left( 1-2\sigma \right) }{E}F_{0}\cos \left( \omega t\right) \frac{%
k_{\perp }}{k^{2}}e^{-%
{\frac14}%
k_{\perp }^{2}r_{_{0}}^{2}},  \label{Teta(k,t)}
\end{equation}
with $k^{2}=k_{x}^{2}+k_{y}^{2}+k_{z}^{2}$.

In the same way, we analytically extend the temperature perturbation $\delta
T$ in the half space $z\leq 0$ in such a way that $\delta T^{\left( 1\right)
}\left( {\bf r},t\right) $ is an even function of $z$. Using the Fourier
transform of the thermal conductivity equation (\ref{Eq.temperature}) and
the expression (\ref{Teta(k,t)}) of $\Theta ^{\left( 0\right) }$, we find
that $\delta T^{\left( 1\right) }\left[ {\bf k},t\right] $ is equal to 
\begin{equation}
\delta T^{\left( 1\right) }\left[ {\bf k},t\right] =A\left[ {\bf k}\right]
e^{i\omega t}+c.c.,  \label{deltaT(k,t)}
\end{equation}
where the function $A\left[ {\bf k}\right] $ is given by 
\begin{equation}
A\left[ {\bf k}\right] =\frac{2\left( 1+\sigma \right) \alpha T}{\rho C}%
\frac{i\omega k_{\perp }}{k^{2}\left( a^{2}k^{2}+i\omega \right) }F_{0}e^{-%
{\frac14}%
k_{\perp }^{2}r_{_{0}}^{2}}.  \label{Eq_Ak}
\end{equation}

We now determine the thermoelastic dissipation $W_{diss}$. The integral over 
$z$ in Eq. (\ref{Wdissipation}) is limited to the volume of the mirror
(infinite half space $z\geq 0$). Since $\delta T^{\left( 1\right) }\left( 
{\bf r},t\right) $ is an even function of $z$, $W_{diss}$ can be written as 
\begin{equation}
W_{diss}=\frac{\kappa }{2T}\left\langle \int d^{3}r\left( {\bf \nabla }%
\delta T^{\left( 1\right) }\right) ^{2}\right\rangle ,
\end{equation}
where the spatial integration is in the whole space. Using the
Bessel-Perseval relation, we can express the dissipated energy $W_{diss}$ as
a function of the temporal average of $\left| \delta T^{\left( 1\right) }%
\left[ {\bf k},t\right] \right| ^{2}$ which is equal to $2\left| A\left[ 
{\bf k}\right] \right| ^{2}$(Eq. \ref{deltaT(k,t)}): 
\begin{eqnarray}
W_{diss} &=&\frac{\kappa }{2T}\int \frac{d^{3}k}{(2\pi )^{3}}%
k^{2}\left\langle \left| \delta T^{\left( 1\right) }\left[ {\bf k},t\right]
\right| ^{2}\right\rangle  \nonumber \\
&=&\frac{\kappa }{T}\int \frac{d^{3}k}{(2\pi )^{3}}k^{2}\left| A\left[ {\bf k%
}\right] \right| ^{2}.
\end{eqnarray}
Using Eq. (\ref{Eq_Ak}), we finally obtain 
\begin{equation}
\frac{W_{diss}}{F_{0}^{2}}=\frac{4T}{\rho C}\alpha ^{2}\left( 1+\sigma
\right) ^{2}\omega ^{2}I,
\end{equation}
where the integral $I$ is given by 
\begin{equation}
I=\int \frac{d^{3}k}{(2\pi )^{3}}\frac{a^{2}k_{\perp }^{2}}{k^{2}\left(
a^{4}k^{4}+\omega ^{2}\right) }e^{-k_{\perp }^{2}r_{0}^{2}/2}.
\end{equation}
This expression allows to determine the spectral density $S_{\widehat{u}}%
\left[ \omega \right] $ of the displacement $\widehat{u}$ from the
fluctuation-dissipation theorem (Eq. \ref{FluctuationDissipation}). One gets
the result given in the text by Eq. (\ref{SuOmega1}): 
\begin{equation}
S_{\widehat{u}}\left[ \omega \right] =32\alpha ^{2}\left( 1+\sigma \right)
^{2}\frac{k_{B}T^{2}}{\rho C}I.
\end{equation}

\end{document}